# Quantum Theory: a Pragmatist Approach

*Richard Healey Philosophy Department, University of Arizona rhealey@email.arizona.edu*

While its applications have made quantum theory arguably the most successful theory in physics, its interpretation continues to be the subject of lively debate within the community of physicists and philosophers concerned with conceptual foundations. This situation poses a problem for a pragmatist for whom meaning derives from use. While disputes about how to *use* quantum theory have arisen from time to time, they have typically been quickly resolved, and consensus reached, within the relevant scientific sub-community. Yet rival accounts of the *meaning* of quantum theory continue to proliferate[1]. In this article I offer a diagnosis of this situation and outline a pragmatist solution to the problem it poses, leaving further details for subsequent articles.

**1. Introduction.** What is it to interpret quantum theory? Addressing this question, van Fraassen (1991) characterized the interpretative task as an attempt to say: "What is really going on, according to this theory?' and "How could the world possibly be how this theory says it is?" This ties interpretation directly to representation: it assumes that the theory offers representations and/or descriptions of the physical world. In (Healey 1989, 6) I expressed sympathy for such a tie as follows: "I should like to add...that a satisfactory interpretation of quantum mechanics should make it clear what the world would be like if quantum mechanics were true." But I continued by noting that it would be inappropriate to criticize a proposed interpretation solely on the grounds that it does not meet this constraint. A theory may further the goals of physics without itself offering novel representations or descriptions of physical reality. If quantum theory is such a theory, then we need an account of how and why it is able to achieve its enormous success. To provide such an account is to offer an interpretation of quantum theory. That is what I set out to do here.

The claim that quantum theory does not itself offer novel depictions of reality may strike some readers as obviously false. What could be more novel than representing the state of a system by a mathematical object such as a wave-function, state vector or density operator, especially when this may represent it as in a superposition, or as entangled with other systems? This is surely quantum theory's distinctive way of describing physical systems, whether or not the description is complete. But there has long been a rival view according to which quantum states convey knowledge or information concerning a system or ensemble without describing its physical condition. I shall elaborate a version of this view that assigns a two-fold role to the quantum state. It plays its primary role in the algorithm provided by the Born Rule for generating quantum probabilities. The quantum state's secondary role cannot be so simply described, but here is the general idea. Any application of quantum theory involves claims describing a physical situation[2]. For example, in an application of the theory to predict or explain results of a contemporary two-slit interference experiment involving detection of individual particles, some claims will describe the apparatus, while others will describe the results of the experiment. But while claims concerning where individual particles are detected contributing to the interference pattern are considered permissible (and even essential), claims about which slit each particle went through are typically alleged to be "meaningless"[3]. The secondary role of the quantum state is to offer guidance on the legitimacy and limitations of descriptive claims about a physical situation. The key idea here is that even assuming unitary evolution of the quantum state of system and environment, delocalization of system state coherence into the environment will typically (though not always) render descriptive claims about experimental results and the condition of apparatus and other macroscopic objects beyond reproach.

[1] In addition to multiple variants of the Copenhagen interpretation, we now have Everettian interpretations of several kinds (many worlds, many minds, ...), the existential interpretation, the transactional interpretation, decoherent histories interpretations, relational interpretations, modal interpretations, de Broglie-Bohm interpretations, quantum Bayesian interpretations, etc.

[2] This may be a situation of a certain type, or a token of that type.

I call this interpretation pragmatist for several reasons. First it takes the uses and applications of the theory to have explanatory priority over its representational capacities. In his trenchant critique of contemporary formulations of quantum mechanics, Bell (1987, 125) unfavorably compared the theory so formulated to classical mechanics: "Of course, it is true that also in classical mechanics any isolation of a system from the world as a whole involves approximation. But at least one can envisage an accurate theory, of the universe, to which the restricted account is an approximation." I doubt that one *can* envisage a detailed and accurate representation of the universe within classical mechanics. Obtaining and using a complete and accurate mathematical model of the universe within classical mechanics would vastly exceed the combined observational and cognitive capacities of humanity or any other physically realizable community of agents, while only in use would such a mathematical model represent anything. But it does not constitute a criticism of a formulation of quantum theory that within it one cannot envisage a complete and accurate representation of the universe, since no successful use or application of quantum theory to cosmology or anywhere else requires that one be able to do so.

The second pragmatist motif concerns the interpretation of quantum probabilities, which are taken to be neither subjective nor straightforwardly objective, but function as a source of authoritative advice to an agent on what to expect and so how to act in specific physical situations. Probabilities derived from the Born Rule do not describe statistical frequencies, even in ideal infinite ensembles: nor do they describe objective chances of individual events. But to accept quantum theory is to commit oneself to apportioning one's partial beliefs in accordance with the probabilities generated by the Born Rule as applied to a quantum state appropriate to one's physical circumstances. I begin to spell this out in more detail in section 2 below.

While not itself issuing descriptive claims about physical reality, quantum theory does advise an agent on the scope and limitations of descriptive claims it may make in a given situation.[4] The advice does not consist in declaring some such claims simply meaningless and so impermissible while others are meaningful and legitimate. Instead the theory places limitations on the inferential power of claims pertaining to the physical situation in which the agent finds itself, or which it represents itself as occupying. Now it is characteristic of pragmatist approaches to meaning to take the content of a descriptive claim to derive ultimately from its inferential relations to other claims and commitments rather than from how it corresponds to the reality it purports to represent.[5] Accepting quantum theory means following its advice to limit the inferential power associated with descriptive claims that may be appropriate in a specific physical situation. So the theory modifies the content of those claims.

[3] Bohmians, of course, reject this allegation. But even they will admit that attempts to verify such claims are either incompatible with appearance of the unaltered interference pattern or themselves rest on equally unverifiable assumptions.

[4] With occasional exceptions for stylistic reasons, I refer to a generalized agent by the pronoun 'it' throughout as a way of emphasizing that users of quantum theory need be neither human nor even conscious, however unlikely the present prospect of nonhuman users.

Bell (1987, 41) introduced the term 'beable' (in contrast to quantum theory's 'observable') to apply to things "which can be described in 'classical terms', because they are there. The beables must include the settings of switches and knobs on experimental equipment, the current in coils, and the readings of instruments." (53) He emphasized that by 'classical terms' he (following Bohr) "refers simply to the familiar language of everyday affairs, including laboratory procedures, in which objective properties—*beables*—are assigned to objects." (41) This at least suggests that a claim about current (for example) derives its content in part from the primitive semantic fact that 'current' refers to (the value of the) current, taken to be intelligible independently of one's disposition to countenance (defeasible) inferences involving this claim (such as the inference that the current consists in the motion of tiny electrically charged particles through a metal, or that it has a source if the current is not zero). More importantly, it assumes that acceptance of quantum theory can in no way modify the content of a claim about beables. But a pragmatist may question that assumption. In section 3 I will offer an alternative account of judgments an agent using quantum theory may make about its physical situation that allows for modification of the content even of claims about its macroscopic environment, including the readings of instruments.

**2. The Objectivity of Quantum Probabilities.** Any attempt to understand quantum theory must address the significance of probabilities derived from the Born Rule, which I write as follows

$$\text{prob}_\rho(A\varepsilon\Delta) = \text{Tr}(\rho \mathbf{P}^{A}[\Delta]) \qquad \text{(Born Rule)}$$

where $A$ is a dynamical variable (an "observable") pertaining to a system $s$, ρ represents a quantum state of that system by a density operator on a Hilbert space $\mathsf{H}_s$ , Δ is a Borel set of real numbers (so $A\varepsilon\Delta$ states that the value of $A$ lies in Δ), and $\mathbf{P}^{A}[\Delta]$ is the value for Δ of the projection-valued measure defined by the unique self-adjoint operator on $\mathsf{H}_s$ corresponding to $A$. Born probabilities yielded by systems' quantum states are the key to successful applications of quantum theory to explain and predict natural phenomena involving them. If one denies that the quantum state describes or represents the physical properties or relations of any system or ensemble of systems, then its main job is simply to yield these probabilities. But what kind of probabilities are these, and what, exactly, are they probabilities of?

If one clear conclusion has been established by foundational work, it is that not every probability derivable by applying the Born rule to a system with quantum state ρ can be taken as a quantitative measure of ignorance or uncertainty of the real-numbered value of a dynamical variable on that system. Born probabilities are not analogous to probabilities in classical

[5] Here is how Brandom (2000, 18) characterizes his pragmatist approach to the conceptual content of a descriptive judgment:

> Pragmatism about the conceptual seeks to understand what it is explicitly to *say* or *think* that something is the case in terms of what one must implicitly know *how* (be able) to *do*. That the relevant sort of doing is a constellation of asserting and inferring, making claims and giving and asking for reasons for them, is the essence of rationalist or inferentialist pragmatism about the conceptual.

statistical mechanics in that they cannot be jointly represented on any classical phase space: quantum observables are not random variables on a common probability space.[6] However, as I expressed it the Born rule specifies, for any state ρ, a probability for each sentence of the form *S*: The value of *A* lies in Δ. The traditional way to resolve the resulting tension is to take each instantiation of the Born rule to observable *A* to be (perhaps implicitly) conditional on *measurement* of *A*, and to assume or postulate that only observables represented by commuting operators can be measured together.[7] Whether this resolution is satisfactory has been the topic of a heated debate.[8] I will address aspects of this in the next section, which offers an account of what the Born probabilities are probabilities of. But whatever they concern, what kind of probabilities are these?

It is common to classify an interpretation of probability as either objective or subjective. Accounts of probability in terms of frequency, propensity or single-case chance count as objective, while the personalist Bayesian interpretation counts as subjective. But then how should one classify classical (Laplacean), logical and "objective" Bayesian notions of probability? It will be best to leave the tricky issue of objectivity aside for a while, so I begin instead by classifying accounts of probability on the basis of their answers to the question "Does a probability judgment function as a description of anything in the natural world?" von Mises's(1922; 2003, 194) answer to this question was clear:

> Probability calculus is part of theoretical physics in the same way as classical mechanics or optics, it is an entirely self-contained theory of certain phenomena . . .

So was Popper's(1967, 32-3)

> In proposing the propensity interpretation I propose to look upon probability statements as statements about some measure of a property (a physical property, comparable to symmetry or asymmetry) of the whole experimental arrangement; a measure, more precisely, of a virtual frequency...

These are expressions of what I will call a *natural property* account of probability.

Accounts of probability as a natural property typically take this to be a property of something in the physical world independent of the epistemic state of anyone making judgments about it. When de Finetti wrote in the preface to his (1974) "PROBABILITY DOES NOT EXIST", this was what he meant to deny. But he wrote elsewhere (1968, 48) that probability

> means degree of belief (as actually held by someone, on the ground of his whole knowledge, experience, information) regarding the truth of a sentence, or event *E* (a fully specified 'single' event or sentence, whose truth or falsity is, for whatever reason, unknown to the person).

and it is at least plausible to suppose that an actual degree of belief is a natural property of the person holding it. If so, even the arch subjectivist de Finetti here adopts a natural property account of probability! Of course, he would insist that different persons may, and often do, hold different beliefs, which makes probability personalist—varying from person to person—and to that extent subjective.

---

[6] Key results are due to Gleason(1957), Bell(1965), Kochen and Specker(1966). The literature now contains many extensions and simplifications, such as Mermin(1990).

[7] An alternative is to privilege a single observable (or commuting family) and take the Born rule as a measure of uncertainty only of the values of privileged observables. Applications of the Born rule to underprivileged observables (conditional on measurement) must then be justified by saying how they are indirectly measured by directly measuring a privileged observable.

[8] See especially Bell(1990), Peierls(1991), Mermin(2006).

On other "subjectivist" views, an agent's degrees of belief count as probabilities only in so far as its overall epistemic state meets a normative constraint of *coherence*,[9] since otherwise these partial beliefs will not satisfy Kolmogorov's(1933) axioms defining probability mathematically as a finitely additive, unit-normed, non-negative function on a field of sets. Ramsey(1926), for one, took probability theory as a branch of logic, the logic of partial belief and inconclusive argument.[10] So viewed, probability theory offers an agent prescriptions for adjusting the corpus of its beliefs so that its total epistemic state meets minimal internal standards of rationality—standards that are nevertheless met by the total epistemic state of few if any actual agents. A probability judgment made by an agent then counts as an expression of its partial degree of belief and a commitment to hold its epistemic state to this minimal standard of rationality. Such a probability judgment does not function as a description of the agent's own belief state, and is certainly not a description of a natural property of anything else in the physical world.

On the present approach, quantum probabilities given by the Born rule do not describe any natural property of the system or systems to which they pertain, or of any other physical system or situation: nor is it their function to describe any actual agent's state of belief, knowledge or information. Their function is to offer advice to any actual or hypothetical agent on the extent of its commitment to claims expressible by sentences of the form *S*: The value of *A* on *s* lies in Δ—roughly, what degree of belief or credence to attach to such a claim.

Consider as an example the experiment of Tonomura *et al.*(1989) in which the positions of electrons are detected in an electron-biprism version of the two-slit experiment by the discrete, localized flashes they produce on a sensitive screen as each makes its contribution to the classic interference pattern. Before running the experiment, an experimenter does not know whether the following statement $S_{34}$ is true: The position of the 34th electron to contribute to the interference pattern is on the left hand side of the screen. Afterwards he or she can check its truth-value by watching the video that recorded each individual flash as it occurred. Assuming the experiment has been correctly set up so that the whole apparatus up to and including the screen has exact left-right symmetry, the Born probability of $S_{34}$ will be ½. Beforehand, an experimenter who makes this assumption and accepts quantum theory should therefore believe $S_{34}$ to the same degree that she disbelieves it. This partial belief will dispose the experimenter to behave in ways she would not have behaved if she had taken the Born probability of $S_{34}$ to be .99: she may accept a bet on $S_{34}$ she would have declined, or she may simply decide it is not necessary to readjust the apparatus to get a more symmetric interference pattern.

This is how quantum probabilities serve as what Bishop Butler famously called the very guide of life. It is quantum theory's great achievement to have made available such a wonderfully reliable guide of such extraordinarily wide applicability. But note that to be guided by quantum theory in this way one needs to know more than just the Born rule—one needs to know what system or type of system to apply it to, and what quantum state to assign to that system. This kind of "know how" is a prerequisite for the successful application of any physical theory, classical as well as quantum. The success of quantum theory is due in large part to the hard-won acquisition of this kind of knowledge by physicists, which has often required great originality and ingenuity. Once acquired, much of it can be conveyed to others and taught to students, although applying Born probabilities to novel situations remains a skill that cannot be mastered by rote learning. But the important point is that knowing what quantum state to assign to a system in order to apply the Born rule constitutes objective knowledge, tacit or otherwise.[11]

---

[9] Dutch book arguments seek to justify coherence constraints by appeal to rationality conditions on betting behavior. Representation theorems justify them by appeal to rationality conditions on preferences. In each case, these conditions are grounded on internalist norms rather than on external states of affairs in the natural world relevant to the content of the beliefs.

[10] But note that Ramsey himself there warned against taking his conclusions to prejudge the meaning of probability in physics.

While there are disputes about what quantum state to assign in a particular situation, these are typically resolved by the same kind of debates within the relevant scientific community as those surrounding the establishment of a novel biochemical structure. Exceptions to this generalization will be discussed in section 4. They are important because they illustrate respects in which quantum states, unlike classical states, are relational. But not everything that is relational is subjective, as debtors and widows are only too aware. What quantum state to ascribe to a system is not at the whim of each agent's subjective beliefs, and nor are the Born probabilities consequent on this ascription. There are at least three reasons why these probabilities are objective.

(1) There is widely shared agreement on them within the scientific community

(2) A norm is operative within that community requiring resolution of any residual disagreements

(3) This norm is not arbitrary but derives directly from the scientific aims of prediction, control and explanation of natural phenomena.

(Details of the derivation must here remain a matter for further investigation.)

Each of these is a reason both for why the Born rule is an essential part of the objective content of quantum theory and for why quantum state ascriptions and the consequent Born probabilities are themselves objective.

After formulating a minimalist account of truth, Wright(1992) presents considerations that may incline one to deploy a richer notion of truth as correspondence to objective reality in some domain. By applying these considerations to ascriptions of quantum probabilities we can further articulate the sense in which these are objective even though they do not describe any natural property of or associated with quantum systems.

The first consideration stems from the Cognitive Command constraint, of which this is an abbreviated version of Wright's first "extremely rough" formulation:

> A discourse exhibits Cognitive Command if and only if it is *a priori* that differences of opinion arising within it can be satisfactorily explained only in terms of "divergent input", "unsuitable conditions", or "malfunction". (92-3)

He gives this later formulation to qualify and simplify his first formulation:

> It is *a priori* that differences of opinion formulated within the discourse, unless excusable as a result of vagueness in a disputed statement, or in the standards of acceptability, or variation in personal evidence thresholds, so to speak, will involve something which may properly be regarded as a cognitive shortcoming. (144)

[11] See Polanyi(1962).

What is the idea of the Cognitive Command constraint, and what motivates it? Here is what Wright says:

> The formulation offered is an attempt to crystallise a very basic idea we have about objectivity: that, where we deal in a purely cognitive way with objective matters, the opinions which we form are in no sense optional or variable as a function of permissible idiosyncracy, but are *commanded* of us—that there will be a robust sense in which a particular point of view *ought* to be held, and a failure to hold a particular point of view can be understood only as a rational/cognitive failure. (146)

This nicely captures the sense in which I claim quantum probabilities are objective. Indeed, one can locate a twofold source of the command in this case. Quantum theory itself commands that quantum probabilities conform to the Born rule: the community's collective evidence-based judgment commands use of a particular quantum state in the Born rule. Neither command is arbitrary: the authority in each case rests ultimately on experimental and observational results and collective judgment of their evidential bearing. But Wright continues

> It is tempting to say that this just is, primitively, what is involved in thinking of a subject matter as purely objective, and of our mode of interaction with it as purely cognitive; and that the Cognitive Command constraint, as formulated, is merely what results when the basic idea is qualified to accommodate various germane kinds of vagueness. ... But...the truth is that the constraint does not reflect a wholly primitive characteristic of the notions of objectivity and cognitive engagement but derives its appeal, at least in part, from a truism to do with the idea of *representation*. For to think of oneself as functioning in purely cognitive mode, as it were, is, when the products of that function are beliefs, to think of oneself as functioning in representational mode...(*ibid*.)

Reflection on the epistemic function of probability ascriptions should prompt one to question this "truism". For while one is clearly functioning in cognitive mode when assessing probabilities, the products of that function are *partial* beliefs, and while each of these does indeed have some kind of representational content, only in the case of derivative, higher-order applications does this concern *probabilities*—in the fundamental situation, the content of each partial belief represents a possible state of the world free of any natural "probability properties".[12] The constitutive function of quantum probability statements is not to represent certain probabilistic aspects of the physical world, but to guide agents in forming appropriate partial beliefs about non-probabilistic aspects of the physical world. The intuition Wright takes his Cognitive Command constraint to express survives undercutting of any possible justification by appeal to alleged truisms about representation.

Wright introduces a second consideration favoring deployment of a richer notion of truth as correspondence to objective reality in some domain.

> Let the *width of cosmological role* of the subject matter of a discourse be measured by the extent to which citing the kinds of states of affairs with which it deals is potentially contributive to the explanation of things *other than*, or *other than via*, our being in attitudinal states which take such states of affairs as object... . The crucial question is... what else there is, other than our beliefs, of which the citation of such states of affairs can

[12] While each partial belief formed in accordance with the Born probability rule does concern the physical world, the next section explains why it is best not to think of this in straightforwardly representational terms.

> feature in.. explanations. (196-7)

Ascriptions of quantum probability have very narrow cosmological role. While they may and do play a role in a huge variety of applications of quantum theory in prediction as well as explanation, in each case the contribution of quantum probabilities is indeed *via* an agent's being in attitudinal states which take quantum probabilities as object. Someone may object that it is a basic role of quantum probabilities to explain frequencies observed, say, in experimental tests of Bell inequalities. But of course a claim about frequencies follows from a claim about probabilities *only with a certain probability*: so a judgment that an observed frequency is explained by a quantum probability itself proceeds *via* an agent's being in attitudinal states which take quantum probabilities as object.

The narrow cosmological role of quantum probability statements provides further support for the conclusion that these do not represent natural properties. But it does nothing to undermine the *objectivity* of quantum probabilities.

Despite his avowed subjectivism about probability, David Lewis(1980) undertook to offer a subjectivist's guide to a kind of objective probability he called chance.

> Along with subjective credence we should believe also in objective chance. The practice and the analysis of science require both concepts. Neither can replace the other. Among the propositions that deserve our credence we find, for instance, the proposition that (as a matter of contingent fact about our world) any tritium atom that now exists has a certain chance of decaying within a year. (1986, 83)

I will explain in section 4 why quantum probabilities should not be taken to have all the features Lewis attributed to chance. But Lewis was right to believe that objective probabilities figure in science, and that these include quantum probabilities. This makes it particularly interesting that he took a single principle to capture all we know about chance, namely the Principal Principle, whose initial statement was as follows:

> Let $C$ be any initially reasonable credence function. Let $t$ be any time. Let $x$ be any real number in the unit interval. Let $X$ be the proposition that the chance, at time $t$, of $A$'s holding equals $x$. Let $E$ be any proposition compatible with $X$ that is admissible at time $t$. Then $C(A/XE)=x$. (1986, 87)

If the only thing we know about objective probabilities is that they command an agent to adjust its credences (partial beliefs) so they equal the corresponding objective probabilities, then it is not surprising that they carry so little explanatory weight. The "thinness" of Lewis's account of probability as it occurs within physics reinforces the application of Wright's point—that narrowness of cosmological role is convincing evidence against a representational view of quantum probability as a natural property of (something in) the world. But this in no way undermines Lewis's claim to be offering an account of *objective* probability.

**3. How quantum theory limits description of physical reality.** In their famous EPR(1935), Einstein, Podolsky and Rosen assumed that quantum theory has descriptive resources but argued that these do not permit a *complete* description of physical reality. Specifically, their argument set out to show that the wave-function fails to give a complete description of the state of an individual system. But they took it for granted that the wave-function's role was indeed to describe physical reality, however incompletely: Einstein(1949; 82-7, 682) suggested that the wave-function should be taken incompletely to describe a statistical ensemble of similarly prepared systems.

On the pragmatist approach I am presenting here, the quantum state does not itself purport to describe physical reality at all—not even incompletely. But in addition to its role in generating quantum probabilities *via* the Born rule, it has an important secondary role in licensing limited claims about physical reality by an agent applying quantum theory.

One can appreciate the need for such a role only after one has abandoned the idea that quantum theory itself makes available new descriptive or representational resources, either in the form of quantum state ascriptions or in some other way (perhaps by allowing dynamical variables to take on operator-values—$q$-numbers—instead of, or as well as, real numbers—$c$-numbers). Rather than thinking of quantum theory as providing distinctively new ways of describing or representing physical reality, focus instead on its effect on *non-quantum* descriptions and representations.

It is tempting to refer to such non-quantum descriptions and representations as *classical*, following Bohr and others. But there are at least two reasons not to yield to this temptation. First, it encourages the mistaken thought that any use of such a description carries with it the full content of classical physics, including dynamical laws such as those of Newton and Maxwell. More importantly, it tends unduly to limit the scope of non-quantum descriptions to exclude what Bell called "the familiar language of everyday affairs, including laboratory procedures" as well as descriptions made available by further advances in non-quantum physics (for example, possible modifications of classical relativity to secure empirical adequacy in light of new observations attributed to high-energy cosmic rays or so-called dark matter). The scope of non-quantum descriptions is very wide: indeed, if quantum theory itself provides no resources for describing or representing physical reality, then *all* present and future ways of describing or representing it will be non-quantum.

It is critical for the present approach to have available non-quantum descriptions of outcomes of quantum measurements. Call a claim expressed by a sentence of the form $S$: "The value of $A$ on $s$ lies in $\Delta$" a non-quantum magnitude claim (NQMC). If one could not express the result of such a measurement in a NQMC, then the Born rule could acquire no empirical support from measurements and we should have little or no reason to believe quantum theory. Quantum theory itself does not imply sentences of the form $S$ on the present approach, even in a case in which the Born rule assigns $S$ probability 1. But physicists make claims using such sentences (or their equivalents) all the time, when describing the results of quantum measurements and in many other circumstances (e.g. in describing the operation of particle accelerators and nuclear reactors, as well as the position of ions in a crystal and the velocity and polarization of photons propagating through an optical fiber delay). How can one reconcile this practice with their acceptance of quantum theory, in light of the no-go results mentioned in section 2 (see footnote 6)?

Answering this question will require excursions into pragmatist philosophy as well as the quantum physics of decoherence. Claims of environmentally-induced decoherence, to solve the quantum measurement problem and explain the emergence of classical behavior of macroscopic objects, are now widely (and wisely) regarded with suspicion.[13] But it is hard to dismiss the thought that decoherence has *some* important role to play in resolving interpretational problems of quantum theory. As Bacciagaluppi(2003/7) and Schlosshauer(2007) explain, decoherence plays different roles within different attempted interpretations of the theory. So after a sketch of relevant quantum physics of decoherence, my main task here will be carefully to explain how a pragmatist can use this to explain when and how quantum theory itself can license the kind of non-quantum descriptive claims physicists do, and must, make in order successfully to apply quantum theory.

[13] See, for example, Bub(1997), Healey(1998), Adler(2003), Janssen(2008).

The basic idea of environmentally-induced delocalization of coherence is well known.[14] It may be illustrated by this toy model. Given an arbitrary superposed pure quantum state of a system *s* interacting with a system *s′* in an appropriate initial quantum state $|b_0\rangle$ , there are Hamiltonians on the tensor product Hilbert space $\mathsf{H}_s \otimes \mathsf{H}_{s'}$ that will induce the following unitary evolution of the total quantum state of *s*+*s*□ :

$$\sum_i c_i |a_i\rangle|b_0\rangle \rightarrow \sum_i c_i |a_i\rangle|b_i\rangle \qquad (2)$$

for some complete orthonormal basis $|a_i\rangle$ of $\mathsf{H}_s$. The resulting quantum state of *s* is then given by partial tracing over $\mathsf{H}_{s'}$ as $\rho_s = \sum_i |c_i|^2 |a_i\rangle\langle a_i|$, which contains no terms diagonal in the preferred $|a_i\rangle$ basis defined by the Hamiltonian for this interaction. Thinking of *s′* as the environment of *s*, such an interaction with its environment has delocalized the coherence of *s*'s initial state into the more inclusive system *s*+*s′* (which in this case remains pure): every Born probability for an observable on *s* alone equals the weighted average (with weights $|c_i|^2$ ) of Born probabilities of all states $|a_i\rangle$. Following such an interaction, *s* will display none of the interference characteristic of quantum mechanical superpositions. To observe any interference it would be necessary to perform an appropriate *joint* measurement involving *both* of *s* and *s′*.

Environmentally induced decoherence has only relatively recently become the subject of experimental investigation. One particularly revealing set of experiments studies interference phenomena involving large molecules including fullerenes ($C_{60}$ and $C_{70}$ molecules). Hackermüller *et al*.(2004) investigated the effects of increased temperature in matter wave interferometer experiments in which $C_{70}$ molecules lose their quantum behavior by thermal emission of radiation. They prepared a beam of $C_{70}$ molecules of well-defined velocity, passed them through two gratings of a Talbot-Laue interferometer in a high vacuum, and detected those that passed through a third movable grating set at the appropriate Talbot distance and used as a scanning mask, by ionizing them and collecting the ions at a detector. Each molecule is sufficiently large and complex to be assigned a temperature as it stores a considerable amount of energy in its internal degrees of freedom. Interaction with the electromagnetic vacuum may result in emission of photons with an intensity and frequency that increases as the internal temperature is raised. These photons may be considered the environment of the molecule. Entanglement between such photon states and the state of the emitting molecule tends to induce environmental decoherence.

Hackermüller *et al*.(2004) present a theoretical model of this decoherence that fits their observations quite well, as the observed interference dies away when the molecules' temperature is raised from 1000°K to 3000°K. This model bears an interesting correspondence to more informal discussions of how the possibility of observing through which slit a particle passed will prevent observation of any consequent interference pattern.[15] Such discussions often focus on particular methods for trying to observe through which slit each particle passes, and proceed to argue that quantum features of the required apparatus necessitate a trade-off between success in this attempt and success in obtaining any resulting interference pattern. Following Heisenberg(1930), one often considers shining light on the particles and collecting reflected light in a microscope focused on them as they pass the slits. In order to tell through which slit a particle goes one would need to use a microscope capable of resolving distances at least as small as the slit separation. Now the resolving power of a microscope is limited by the wavelength of light used: better resolving power requires light of shorter wavelength. However, photons of light of short enough wavelength would have such a large momentum as to disturb the particle and effectively to destroy the interference pattern. Even though no observation of the positions of $C_{70}$ molecules as they pass through the apparatus is contemplated in these experiments, and no light is shone on them, the theoretical model of decoherence shows that the possibility of photon *emission* of short enough wavelength to make it *possible* to determine through which slit each molecule goes is enough effectively to destroy the interference pattern. Moreover, the detailed form of the quantum state of the fullerene (expressed in the off-diagonal elements of the fullerene center-of-mass position density operator) describes the diffraction limitation of a hypothetical microscope used to obtain which-path information on the molecules.

---

[14] See, for example, Joos *et al*.(2003), Schlosshauer(2007), Zurek(2003, 2009).

[15] See, for example, Feynman(1963, vol. III, 1.8-9).

The phrase "which-path information" (or „welcher-Weg-Information") that occurs repeatedly in Hackermüller *et al.*(2004) and many other experimental as well as theoretical treatments of quantum interference is puzzling but highly suggestive. I shall pursue the suggestion after discussing another related experiment recently conducted in the same laboratory in Vienna.

Juffman *et al.*(2009) prepared a beam of $C_{60}$ molecules with well-defined velocity, passed them through two gratings of a Talbot-Laue interferometer in a high vacuum, and collected them on a carefully prepared silicon surface placed at the Talbot distance. They then moved the silicon about a meter into a second high vacuum chamber and scanned the surface with a scanning tunneling electron microscope (STEM) capable of imaging individual atoms on the surface of the silicon. After running the microscope over a square area of approximately $2\mu m^2$ they were able to produce an image of some one to two thousand $C_{60}$ molecules forming an interference pattern.[16] They reported that the surface binding of the fullerenes was so strong that they could not observe any clustering, even over two weeks. Clearly they felt no compunction in attributing very well defined, stable, positions to the molecules on the silicon surface, and even recommended developing this experiment into a technique for controlled deposition for nano-technological applications.

Together, these experiments illustrate three different scenarios in which one may contemplate making a claim about the position of an individual fullerene molecule involved in a quantum interference experiment. By reflecting on the inferential commitments entered into by one who makes such a claim, we shall be able to gain a better appreciation of the significance of judgments expressed in NQMCs of the form *S*: 'The value of *A* on *s* lies in Δ', beginning with the case in which *s* is an individual fullerene molecule, *A* is the horizontal distance *x* (in nanometers) of its center of mass from a reference point in the plane of the vertically oriented gratings and Δ is an interval of real numbers. To repeat, while quantum theory itself does not imply sentences of the form *S*, they play an essential role in any application of quantum theory.

[16] While initial scanning was performed at room temperature, cooling to low temperatures reduced thermal motion sufficiently to image the internal structure of individual molecules.

After $C_{60}$ molecule *s* has been deposited on the silicon substrate in the experiment of Juffman *et al.*(2009) and imaged by the STEM, their figures 2 and 3 (together with the surrounding discussion) illustrate that some claim of the form $S_x$: 'The position *x* of *s* is $x_s \pm \varepsilon$' for some value of $\varepsilon<5$nm is warranted. The warrant derives substantially from the reliability of the image-forming process, importantly including the (quantum!) theory and practice underlying the successful operation of the STEM used to produce it. But there is a prior issue: given that a $C_{60}$ molecule may itself be treated as a quantum system, how and why is one entitled to attribute to *s* a definite, stable position in the first place?

It is in answering this question that it is appropriate to appeal to environmental decoherence. While it may be difficult to formulate and solve the Schrödinger equation for a realistic many-body quantum interaction that binds *s* to the silicon surface, it is clear that this will rapidly and strongly couple *s* to an environment of an exponentially increasing number of degrees of freedom, involving the entire silicon crystal and light reflection from its surface, thermal radiation interacting with phonons in the crystal, vibrations and thermal motion of the supporting structure of the crystal, and eventually the entire laboratory and beyond. Examination of the properties of analytically and computationally solvable models of decoherence in simpler systems justifies one in concluding with a high degree of confidence that the center-of-mass state of *s* alone will extremely quickly become, and remain indefinitely in the absence of external disturbances, very close to diagonal in a preferred "position basis" of states, each close to a delta function of position.

It does not follow that some statement of the form $S_x$ is true. On the present approach, no analysis of a decoherence interaction to show the (approximately) diagonal form of the quantum state of a decohering system ever itself thereby implies any such non-quantum statement. The import of the quantum analysis is more subtle. What decoherence shows in this example is that what an agent may legitimately infer from a claim about the position of *s* of the form $S_x$ is, as it relates to any conceivable goal of that agent, exactly what would follow from the simple truth of $S_x$.[17] This is how the quantum theory of decoherence *licenses* the experimenters in Juffman *et al.*(2009), anyone reading their paper, and indeed any suitably physically situated agent, human, conscious, or neither, to make some such claim. While quantum theory in this way *licenses* many incompatible claims of this form, each ascribing a different value $x_{s1}$, $x_{s2}$ , $x_{s3}$ ,... to *x*, by itself the theory *warrants* an agent in claiming none of them: that requires additional, reliable empirical information of a kind acquired by the skillful use of the STEM used by Juffman *et al.*(2009) to produce data like that displayed in figures 2 and 3. Quantum licensing takes the following form: a quantum state of a system and its environment may be such as to grant an agent permission to issue a judgment of a certain kind concerning that system. Equipped with the necessary permission, the agent may be warranted by its "experience" to issue one rather than another judgment of that kind.

Feynman(1963, vol. III, 1.9) said this about the position of an electron as it passes through an analogous 2-hole interference experiment:

> if one has a piece of apparatus which is capable of determining whether the electrons go through hole 1 or hole 2, then one *can* say it goes through either hole 1 or hole 2. [otherwise] one may *not* say that an electron goes through either hole 1 or hole 2. If one

[17] This claim will be further elaborated and defended in section 5.

*does* say that, and starts to make any deductions from the statement, he will make errors in the analysis. This is the logical tightrope on which we must walk if we wish to describe nature successfully.

Consider instead the status of claims of the form $S_x$ about a fullerene *s* as it passes through the interferometer gratings in either of the two experiments just described. In each diffraction grating in these experiments the slits were regularly spaced at a distance of some hundreds of nanometers. So if one could say $S_{or}$: $Sx_{s1}$ or $Sx_{s2}$ or $Sx_{s3}$ or ... (where $x_{si}$ marks the center of the *i*th slit and ε now corresponds to the width of each slit) then one could say the fullerene goes through slit 1 or slit 2 or slit 3 or ... . But *can* one say $S_{or}$?

In the experiment of Juffman *et al.*(2009) there was no piece of apparatus capable of determining which slit each fullerene goes through. What goes wrong if one says $S_{or}$? Feynman's discussion makes clear the nature of the error he thinks would follow from this claim. Suppose one assumes the *j*th fullerene goes through slit *i*. It would have made no difference to its subsequent behavior if all the other slits had been closed, so it would have contributed to a single slit interference pattern centered on slit *i*. It follows that the total interference pattern will be just a sum of single slit patterns for all *i*, weighted by the number of fullerenes going through each slit. Since the actual interference pattern is quite different, $S_{or}$ has been empirically falsified.

The form of the argument is *reductio ad absurdum*, but as is typical for such arguments it rests on additional premises, any of whose rejection prevents one from drawing the intended conclusion. Bohmians, among others, have principled reasons for denying that the behavior of a fullerene passing through one slit is independent of whether other slits are open or closed: roughly, they take its behavior to be governed by a physically real wave-function that passes through *all* the open slits. Bohmian mechanics shows how to draw many more interesting conclusions of $S_{or}$ consistent with quantum-theoretic predictions, though at the cost of accepting action-at-a-distance.

If one supplements $S_{or}$ with *no* additional premises, one will never risk error in making deductions. This is a trivial consequence of two facts: (i) $S_{or}$ is logically consistent, and (ii) no valid deductive argument can lead from logically consistent premises to a contradictory conclusion. But this response misses Feynman's point, since he is clearly concerned not just with formally valid deductive arguments whose sole premise is $S_{or}$, but with inferences of the kind anyone with a normal understanding of $S_{or}$ will naturally make, for example that:

I. No particle passes through the material (silicon nitride) in which the slits are cut

II. It is possible reliably to observe through which slit each particle passed without altering the interference pattern, or

III. If this is not so, then that can only be because any physical mechanism that permitted reliable observation of through which slit each particle passed would inevitably disturb the particle while doing so

Brandom(2000), following Sellars(1953), calls inferences such as those from $S_{or}$ to I, II, III *material* inferences. Here and in Brandom(1994) he develops what he calls an inferentialist pragmatism about conceptual content. It is a consequence of this kind of pragmatism that the *content* of $S_{or}$ is a function of the material inferences that connect it to other claims and other actions by a claimant or others in the same linguistic community. Accepting quantum theory in no way undermines the inference from $S_{or}$ to I: this remains a legitimate material inference even though it is not formally valid. But however natural inferences to II or III may seem, application of quantum theory shows that both II and III lead to conflict with results of experimental (or at least *Gedankenexperimental*) findings.[18] So while one can *say* $S_{or}$ (*pace* Feynman), the content of that claim must be understood very differently within a community that has accepted a quantum theoretic analysis of the situation to which the claim applies. Given the possibility of confusion provided by so severely weakening the claim, it may be wise to heed Feynman's cautionary advice not to say $S_{or}$ at all in the context of the experiment of Juffman *et al.*(2009).

The status of $S_{or}$ in the experiment of Hackermüller *et al.*(2004) is more complex. In the case of low temperature fullerenes, there is relatively little decoherence of their center of mass motion through the interferometer, so the analysis goes through as for the experiment of Juffman *et al.*(2009). While one *can* say $S_{or}$, it is probably safest not to do so, since anyone who did so would naturally be understood as committed to inferences and other actions they did not intend. Without careful qualification, the weakened content of the claim would make it likely subject to misinterpretation. As the temperature of the fullerenes is increased, the interference contrast decreases. The authors comment

> This is the signature of decoherence due to the enhanced probability for the emission of thermal photons that carry 'which-path' information. ... They transmit (partial) which-path information to the environment, leading to a reduced observability of the fullerene wave nature. ... Around 3,000°K the molecules have a high probability to emit several visible photons yielding sufficient which-path information to effect a complete loss of fringe visibility in our interferometer.

I believe it would misinterpret their use of the phrase 'which-path information' here to take them to presuppose that each fullerene follows a determinate, though unknown, path through the slits, which becomes progressively more open to potential observation as its temperature increases. It is better to regard the *content* of a claim made by a statement like $S_{or}$ as itself a function of temperature, in the following sense: as the temperature is increased from 1000°K to 3000°K, the inferential power of the claim increases accordingly. This is why it becomes more and more appropriate to think and speak of the fullerenes as having a well-defined path through the interferometer as the degree of thermally induced electromagnetic decoherence into their environment increases. But note that on the present inferentialist view of content, this progressive definition of content has no natural limit such that one could say that when this limit is reached a statement like $S_{or}$ is simply true because one has finally succeeded in establishing a kind of natural language-world correspondence relation in virtue of which the statement correctly represents some radically mind- and language-independent state of affairs.

It is important to bear this in mind when reconsidering the role of measurement in quantum theory. I have been careful not to formulate the Born rule narrowly so that it explicitly concerns results of measurements on a quantum system. But it is vital that situations to which the Born rule applies include those in which scientists are warranted in making claims about the values of dynamical variables as a result of performing operations they take to constitute measurements of them. Recall from the introduction how Bell introduced his notion of beables to apply to things

> …which can be described in 'classical terms', because they are there. The beables must include the settings of switches and knobs on experimental equipment, the current in coils, and the readings of instruments.

emphasizing that by 'classical terms' he

> refers simply to the familiar language of everyday affairs, including laboratory procedures, in which objective properties—*beables*—are assigned to objects.

---

[18] Although it would be very difficult to modify the experiment of Juffman *et al.*(2009) to reliably observe through which slit each particle passed on its way to the silicon detector, claims of the form II and III have been experimentally refuted in analogous experiments with fewer "slits". While observing through which slit each particle passed in such an experiment does require some interaction that correlates a particle's state with that of a "probe" system, this need not disturb the *particle's* state. So-called "quantum eraser" experiments demonstrate that even if an interference pattern *is* destroyed by an interaction that disturbs the particles' states, it may be restored by a suitable interaction directly involving neither particle nor "probe" system. Moreover, for each particle, the restoring interaction can be delayed until after the particle is detected: see Walborn *et al.*(2002) for an actual two-slit delayed quantum erasure experiment.

His thought seems to be that at least when it comes to descriptions of experimental equipment and laboratory procedures language must be taken to function in a straightforwardly representational way—as simply saying how things are.

What "is there" in these fullerene experiments? Because of the massive decoherence between large scale features of the macroscopic laboratory apparatus and its environment, quantum theory licenses claims about the settings of switches and knobs on experimental equipment, the macroscopic current in coils, and the readings of instruments. The content of such claims is almost, but not quite, unchanged by acceptance of quantum theory, since the limits the theory places on their inferential power are of no importance for any practical, or even impractical, purpose. Acceptance of quantum theory *does* significantly modify the content of claims about the microscopic currents produced by electrons tunneling from the fullerenes and silicon surface when scanned by the STEM in the experiment of Juffman *et al.*(2009). These currents cannot be said to "be there" in the same robust sense, in so far as each results from a tunneling process that is characteristically non-classical. And, as the preceding discussion made clear, acceptance of quantum theory so significantly limits the content of claims about the position of fullerenes as they pass through the interferometer that anyone making them at best courts confusion. But the efficiency of decohering interactions between a fullerene molecule and atoms of the silicon surface in the experiment of Juffman *et al.*(2009) is such that even though the molecule and atoms are microscopic and the interaction is quantum, a claim that the molecule is there at a specific location on the surface has almost the status of a claim that the entire apparatus is there in the laboratory.

There is no explicit reference to measurement in the published report of either fullerene experiment. Moreover, in applying quantum theory to account for the features of the interference patterns it is not necessary to interpret the Born rule explicitly to concern probabilities of *measured* positions of fullerenes in the pattern: one can take it simply to give probabilities for their positions at the detector. But I think it is clear that deposition of a $C_{60}$ molecule on the silicon substrate in the experiment of Juffman *et al.*(2009) does count as performance of a quantum measurement of the molecule's position. Certainly there is no temptation to say that the molecule has no definite position on the surface until and unless a subsequent observation is carried out using the STEM.

We can now see both why it is natural to formulate the Born rule so that it concerns probabilities *of measurement outcomes* and why the application of that rule is not restricted to measurement contexts. One reason to explicitly mention measurement in a formulation of the Born rule is to stress that the evidence justifying acceptance of quantum theory rests to a large extent on the results of experiments in which observables are measured and the statistical distribution of measurement outcomes compared to those expected on the basis of the Born rule. If the Born rule could not be connected to measurement outcomes in this way, quantum theory would be cut off from its evidential base. But, as we have seen, the link can be preserved by simply assuming that the outcome of a quantum measurement can be expressed in a NQMC, with no mention of any measurement of a kind needed to determine the value of the magnitude in question. This leads to the second, more substantial reason for formulating the Born rule in terms of measurement: the no-go results mentioned in section 2. Not all dynamical variables on a quantum system can consistently be assigned simultaneous real values distributed in accordance with the Born probabilities—not even so as to match just the extremal Born probabilities 0 and 1. But this presents no problems on the present approach, since Born-rule probabilities are well-defined only over claims licensed by quantum theory. According to the quantum theory, interaction of a system with its environment typically induces decoherence in such a way as (approximately) to select a preferred basis of states in the system's Hilbert space. Quantum theory will fully license claims about the real value only of a dynamical variable represented by an operator that is diagonal in a preferred basis: it will grant a slightly less complete license to claims about approximately diagonal observables. All these dynamical variables can consistently be assigned simultaneous real values distributed in accordance with the Born probabilities. So there is no need to formulate the Born rule so that its probabilities concern only measurement outcomes.

**4. The relational nature of quantum states.** There is a sense in which quantum states are relational on the present pragmatist approach. It is important to appreciate their relational character if one is to understand, among other things, the status of von Neumann's "projection postulate" (wave-packet collapse), why violation of Bell-type inequalities poses no threat of non-local action, why quantum probabilities are not simply Lewisian chances, and how to resolve the "paradox" of Wigner's friend. But first it is necessary to say what this relational nature amounts to, and to distinguish it from other senses in which quantum states have been taken to be relational.

Rovelli(1996, 2005) proposed a relational view of quantum states. This maintains the tight connection between a quantum state and the values of dynamical variables in that state that has come to be known as the 'eigenstate-eigenvalue link': dynamical variable $A$ has value $a$ on $s$ if and only if the quantum state of $s$ is an eigenstate, with eigenvalue $a$, of the self-adjoint operator uniquely corresponding to $A$. Rovelli assumes that both quantum state and associated values of dynamical variables describe or represent the physical condition of an individual system $s_1$, but only relative to some *other* physical system $s_2$. This second system will in turn have a quantum state and associated values of dynamical variables relative to $s_1$ (as well as to $s_3$, $s_4$,...): even though Rovelli sometimes calls it "the observer", he stresses that $s_2$ need be neither human, conscious, classically described, macroscopic, nor "special" in any way—it is just some other quantum system. On Rovelli's relational quantum mechanics, a quantum system has a quantum state and associated values of dynamical variables *only* relative to (any) other distinct quantum system, and these relative states will in general differ according to *which* other system one relativizes them to.

Rovelli's relational view of quantum states is quite different from the pragmatist view I am presenting, which begins by dismissing the eigenstate-eigenvalue link as not merely false but arising from confusion between the radically different roles of quantum state ascriptions and claims about the values of dynamical variables in quantum theory. A second basic difference concerns what each view takes quantum states to be relative to. For the pragmatist, a quantum state ascription is not relative to an arbitrary distinct quantum system, but rather to the perspective of an actual or potential *agent*—some physically situated *user* of quantum theory. While every actual physically situated user of quantum theory may be treated as a quantum system (by some user of quantum theory), not every quantum system is a physically situated user of quantum theory.

Fuchs(2010) advocates what he calls a quantum Bayesian approach to quantum theory (QBism), and seeks to explore its connections to pragmatism, among other philosophies. QBism bears many similarities to the present pragmatist approach. QBism also rejects the eigenstate-eigenvalue link as a misconceived attempt to understand quantum states as yielding descriptions of physical reality. A pragmatist will surely endorse Fuchs's(2010) view that quantum theory as a whole is "a users' manual that any agent can pick up and use to help make wise decisions in this world of inherent uncertainty". Moreover, QBism agrees that quantum states are relative, to the extent that different agents can consistently assign different quantum states to the same system. But unlike the present pragmatist approach, QBism is committed to a subjective Bayesian view of probability that denies that quantum probabilities derived from the Born rule can ever be authoritative for a rational agent who accepts quantum theory. The key difference is that while, for the QBist, quantum state ascriptions depend on the *epistemic* state of the agent who ascribes them, on the present pragmatist approach what quantum state is to be ascribed to a system depends only on the *physical circumstances* defining the perspective of the agent (actual or merely hypothetical) that ascribes it. I will spell this out in more detail after pointing to a contrast with a third recent account of the relational nature of quantum states.

Bartlett *et al*.(2006, 2007) take a quantum state ascription to be relative to what they call a reference frame. In (2006) they motivate this as a way of resolving disputes that have arisen among physicists in a variety of contexts as to whether it is correct to assign to a system a quantum state that is a superposition or an incoherent mixture of eigenstates of some observable. The example they focus on is a dispute as to whether the quantum state of a laser operating above threshold is a coherent state (with a definite phase) or a mixture (that may be represented either as a uniform integral over projections onto coherent states of every phase, or alternatively as diagonal in a photon number basis). They offer to resolve this and analogous disputes by supposing that *each* state ascription may be considered equally correct, but relative to a different reference frame—in this case, relative to a different phase reference frame. They take a reference frame to be embodied in some physical object: in the example, some local oscillator could serve as a phase standard. The laser could be consistently ascribed a coherent state relative to a correlated ("implicated") oscillator (such as the main beam in a homodyne detection experiment) and at the same time an incoherent state relative to an uncorrelated ("non-implicated") reference frame (like the beam from an independent laser). In their view

> ...the whole debate presumes that quantum states only contain information about the *intrinsic properties* of a system. We submit that this presumption is mistaken; quantum states also contain information about the *extrinsic properties* of a system, that is, the *relation* of the system to other systems external to it, and whether or not coherences are applicable depends on the external system to which one is comparing. (2006, 28)

A similar analysis would apply to any analogous dispute involving an observable (such as spin-component) which may itself be thought of as relational.

That quantum states are relational in this sense is interesting, especially because of the way it helps resolve disputes about quantum state ascription that may otherwise have been held up as counterexamples to section 2's claim that these are rare and short-lived. But while endorsing the resolution of these disputes offered by Bartlett *et al*.(2006, 2007), I take quantum states to be relational in a way distinct from relativity to reference frames in their sense. Specifically, quantum states are relational because any ascription of a quantum state to a system relates that system to a physically characterized situation that may (but need not) be occupied by a physically situated agent. A system's quantum state is a state appropriate for any agent bearing the relevant kind of physical relation to that system, so the same system may be ascribed different states for different physical agent situations. Note that an agent situation need not actually be occupied by any agent, just as no observer need actually occupy an inertial reference frame, and recall that the term 'agent' is being used very broadly so as to apply to any physically instantiated user of quantum theory, whether human, merely conscious, or neither.

One important aspect of an agent situation is its temporal relation to the time for which a system's quantum state is to be specified. It is by taking careful note of this relation that one can appreciate the significance of discontinuous changes in quantum states on measurement.

Consider the following example of a so-called negative result measurement. Suppose a source produces photons linearly polarized at 45° to the vertical: each such photon is heralded by detection, after passing through a 45° oriented polarizer, of a second photon of an entangled pair produced in parametric down conversion in a suitable nonlinear crystal.[19] One such heralded photon is incident on a polarizing beam-splitter, in the vertical channel of which is located a high-efficiency photon detector. If nothing is detected, the photon is ascribed the horizontal polarization state $|H\rangle$: the measurement has projected its superposed polarization state as follows

$$1/\sqrt{2}\,(\,|H\rangle + |V\rangle\,) \rightarrow |H\rangle \qquad (3)$$

How can such projection be reconciled with the unitary evolution of the combined state of the photon and detector?

The answer to this question is that the quantum state of the photon's polarization is a superposition of horizontal and vertical relative to the situation of an agent prior to the decohering interaction with the detector and its environment, but horizontal relative to the situation of an agent after that interaction. Decoherence involves no violation of unitarity. Instead, it warrants an agent in using the latter quantum state rather than the former to guide its expectations after judging that the detector has failed to detect the photon—a judgment that is licensed by the form of the unitarily evolved joint quantum state (which correlates the preferred "pointer basis" of detector states to horizontally/vertically polarized basis states of the photon) and warranted by its record of the detector's failure to detect the photon. As Zurek(2009) explains, the post-interaction state of the detector and its environment acts as a "witness" of the horizontal polarization quantum state that may be consulted in many independent ways by an agent. Quantum theory cannot explain that such an agent records the detector's failure rather than success—that remains outside its purview. But it can account for the "intrasubjective" concordance of the agent's records as it performs such multiple independent checks on the detector and the photon itself.[20]

[19] Assume, for now, that such a distant measurement on the entangled partner will produce this polarization state in the heralded photon. This assumption will be scrutinized shortly.

A standard objection to the claim that decoherence can account for definite outcomes of quantum measurements may seem to apply also to this answer.[21] Even if interactions rapidly and robustly entangle the joint state of photon and detector with the state of their environment so that their joint quantum state is a mixture of product states with no off-diagonal terms, this remains an improper mixture that cannot be understood to represent an agent's state of ignorance as to their actual joint (pure) product state. There will be some complex collective dynamical variable on the combined photon, detector and environment system whose measurement will almost certainly display statistics distinguishable from those predicted by such a mixture of product states each correlated with a corresponding environment state. By repeatedly measuring the values of this dynamical variable on identically prepared photon+ detector+ initial environment systems, some "super-agent" could verify that their total state evolves unitarily in a way that is inconsistent with the assumption that the photon+detector system is in some (unknown) pure product state.

This objection cannot be turned into a good argument against the reconciliation of unitary evolution and effective "collapse" offered two paragraphs earlier: That reconciliation nowhere assumed that the quantum state of photon+detector was some (unknown) pure product state. Instead, it simply assumed unitary evolution of the total state, including the environment, to show that, conditional on recording the failure of the detector to detect a heralded photon, the Born probability for a recording of any measurement of the polarization of that photon (even including joint measurements with its environment) is exactly as predicted by assignment to it of quantum state $|H\rangle$. This is why *the photon's* quantum state is $|H\rangle$, relative to the situation of an agent in a position to access records of the detector's failure to fire.

So an agent with a record that the detector has not fired after the decohering interaction between photon and detector (together with its environment) is warranted in ascribing polarization state $|H\rangle$ to the photon. But the agent is *not* thereby warranted in inferring that this photon is horizontally polarized—an inference in conformity to the eigenstate-eigenvalue link. All the Born rule authorizes such an agent to do is to adopt a maximal degree of belief (1) that, following a second (ideal) interaction involving that photon that correlates its horizontal/vertical polarization state with the decohered state of some other polarization detector, the agent would record that detector as recording horizontal. At this point the distinction between quantum state and actual polarization may seem unmotivated. Its significance will become clear in section 5, which analyzes descriptions of a system by more than one agent. It is essential to maintain a distinction between relational quantum state ascriptions and non-relational dynamical variable ascriptions in order to ensure that applications of quantum theory carry objective import.

Consider an experiment set up to investigate violations of Bell inequalities in entangled photon pairs in which Bob measures polarization of photon $R$ along axis $b$ while Alice measures polarization of photon $L$ along axis $a$. Suppose Alice, Bob and everyone else in their expert community is warranted in agreeing that the experiments are performed on photon pairs in the

---

[20] I will address the important further issue of concordance among the records of more than one independent agent in section 5.

[21] See, for example, D'Espagnat(1990).

entangled polarization state

$$|\Phi^+\rangle = 1/\sqrt{2} ( |HH\rangle + |VV\rangle ) \qquad (4)$$

Suppose Bob's measurement is concluded at a time $t_b$ before Alice performs her measurement. Bob then records his detector as indicating polarization of $R$ along (rather than orthogonal to) the $b$ axis, and ascribes polarization state $|b\rangle$ to photon $L$ subsequent to $t_b$. This ascription is warranted by considerations parallel to those that warranted ascribing polarization state $|H\rangle$ to the photon discussed in the immediately preceding paragraphs. Before she performs her measurement on $L$, no such considerations warrant Alice in ascribing state $|b\rangle$ to photon $L$. Instead, Alice is warranted in ascribing to $L$ the same maximally-mixed polarization state given by tracing $|\Phi^+\rangle$ over the polarization Hilbert space of $R$ as before Bob's measurement. The quantum state of $L$ relative to Alice's agent situation is different from the quantum state of $L$ relative to Bob's agent situation. This is true whether or not there *are* any agents Alice and/or Bob actually occupying those situations: all that matters is that for a period of time after $t_b$ the R detector has suitably interacted with $R$ and its environment but the $L$ detector has not.

In an experiment like this, very little time will elapse between the $R$ and $L$ detection events. Indeed, in certain experiments the interval separating them is space-like rather than time-like. These provide further illustrations of the relational nature of quantum state ascriptions. Suppose that Alice and Bob are moving so that while Bob represents the interaction of $R$ with his detector to have concluded before the interaction of $L$ with Alice's detector, Alice takes these events to have occurred in the opposite time-order. Alice will then be warranted in ascribing quantum state $|a\rangle$, say, to $R$ subsequent (for her) to the time $t'_a$ she represents as the conclusion of her measurement of $L$. But if $a \neq b$ Bob will never be warranted in ascribing state $|a\rangle$ (or its Lorentz transform) to $R$, and nor will Alice ever be warranted in ascribing state $|b\rangle$ ⟩ (or its Lorentz transform) to $L$. Again, there need be no actual agents Alice and Bob moving in these ways. But note that specification of an agent situation here involves not only specification of a time interval, but also of a frame (inertial or otherwise) with respect to which that interval is defined.

We can now begin to see why quantum probabilities are not simply chances, as Lewis describes them. Section 5 will show how this also helps reconcile quantum theory's violation of Bell inequalities with a physically motivated locality requirement.

Consider the following space-time diagram (in the laboratory frame), in which the diagonal lines mark boundaries of the causal past or future of the space-like separated events $M_A$, $M_B$ at which $L$, $R$ respectively interact appropriately with a polarization detector and its environment:

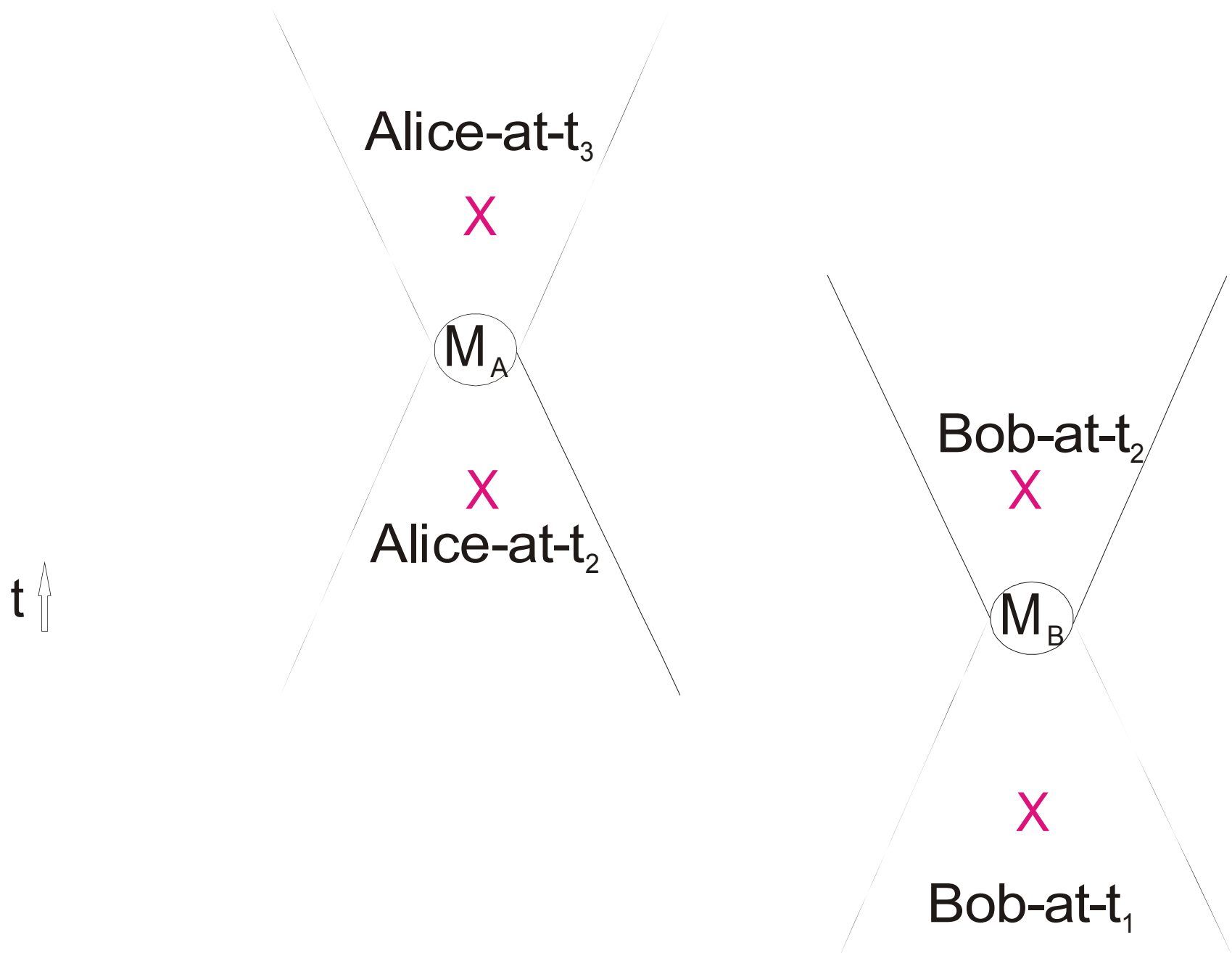

Space-time diagram for non-local correlations

Suppose one asks: What is the probability at $t_1$ that Alice's detector will record *L* with polarization *a*? Since all agree that the quantum state of the pair at $t_1$ is $|\Phi^+\rangle$, the answer ½ follows uniquely by the Born rule. Assume for the moment that *a*=*b*, and consider this question: What is the probability at $t_2$ that Alice's detector will record *L* with polarization *a*? Applying the Born rule to her quantum state for *L* at $t_2$, Alice will give the same answer as before, namely ½. Bob will apply the Born rule to *his* quantum state for *L* at $t_2$ and give the different answer 1. If quantum probabilities obey Lewis's Principal Principle that he takes to capture all we know about chance, then at least one of these answers must be wrong, since that principle presupposes that, while the chance of an event may change as time passes, it is uniquely defined at any given time. But just as quantum state ascriptions are relational, so also are Born probabilities. Relative to Alice's agent situation the probability is ½, relative to Bob's agent situation the probability is 1. Neither probability is subjective, but both are correct. Each is authoritative for any agent that happens to be in the relevant agent situation, and accepting that this is so is a requirement on any agent that accepts quantum theory, whatever the actual situation of that agent.

Notice that adapting the framework of Lewisian chance to relativistic space-time structure by allowing the chance of an event to depend not on some absolute time but rather on the time in any reference frame will not effect a reconciliation between apparently conflicting quantum probabilities Alice and Bob should assign to the same event here. For Bob in his reference frame, there will be a time interval after $M_B$ and before $M_A$ during which he should assign probability 1 to Alice's recording polarization *a* on photon *L*. But this probability assignment could play no role in the decision-making of any agent at rest in Bob's reference frame but within the back light-cone of $M_A$, since the outcome of Bob's measurement lies outside her back light-cone and should be counted by Lewis as inadmissible information for her (assuming no superluminal signaling): the objective probability for her remains ½, as specified by the Born rule applied to her quantum state. The relativization of chance to an arbitrary foliation by space-like Cauchy surfaces would face essentially the same objection: the value of the chance at some point on a Cauchy surface would be irrelevant to the decision-making of an agent at that point. One could try to understand quantum probabilities as Lewisian chances of an event by specifying them only on space-like hypersurfaces restricted to the causal past of that event—the closure of the event's back light-cone. But this proves problematic for a different reason. To take the value of the chance of an outcome of $M_A$ on a space-like hypersurface within the back light-cone of $M_A$ to be given by Alice's Born probabilities is to ignore the relevant information provided by the record of Bob's measurement that is available to Bob at $t_2$: On the other hand taking the value of this chance to be given by Bob's Born probabilities both makes it arbitrary on which space-like hypersurface within $M_A$'s causal past they change and raises the specter of non-local action.

The relational nature of quantum state ascriptions and the consequence that Born probabilities are not simply time-relativized Lewisian chances may be brought home even more forcefully by examining an experiment first proposed by Peres(2000): a version of the experiment has recently been successfully performed. Since the experiment combines two independently interesting quantum phenomena, I begin by discussing the first—entanglement swapping—before introducing the second—delayed choice.

Two agents, Alice and Bob, simultaneously but independently prepare pairs of polarization-entangled photons in (universally agreed) quantum state $|\Psi^-\rangle = 1/\sqrt{2}\,(|HV\rangle - |VH\rangle)$: call Alice's photons 1,2 and Bob's 3,4. Alice measures the polarization of photon 1 along axis *a*, while Bob measures the polarization of photon 4 along axis *b*. Photons 2 and 3 are passed through optical fiber delays before each is incident on a beam splitter as indicated in the accompanying space-time sketch. A third agent, Victor, then measures the polarization along axis *H* of any photons emerging to the left of the beam splitter, and the polarization along axis *H* of any photons emerging to the right of the beam splitter. After allowing for the delay, careful timing permits recording of fourfold coincidence counts from detection of all four photons from two pairs simultaneously prepared by Alice and Bob.

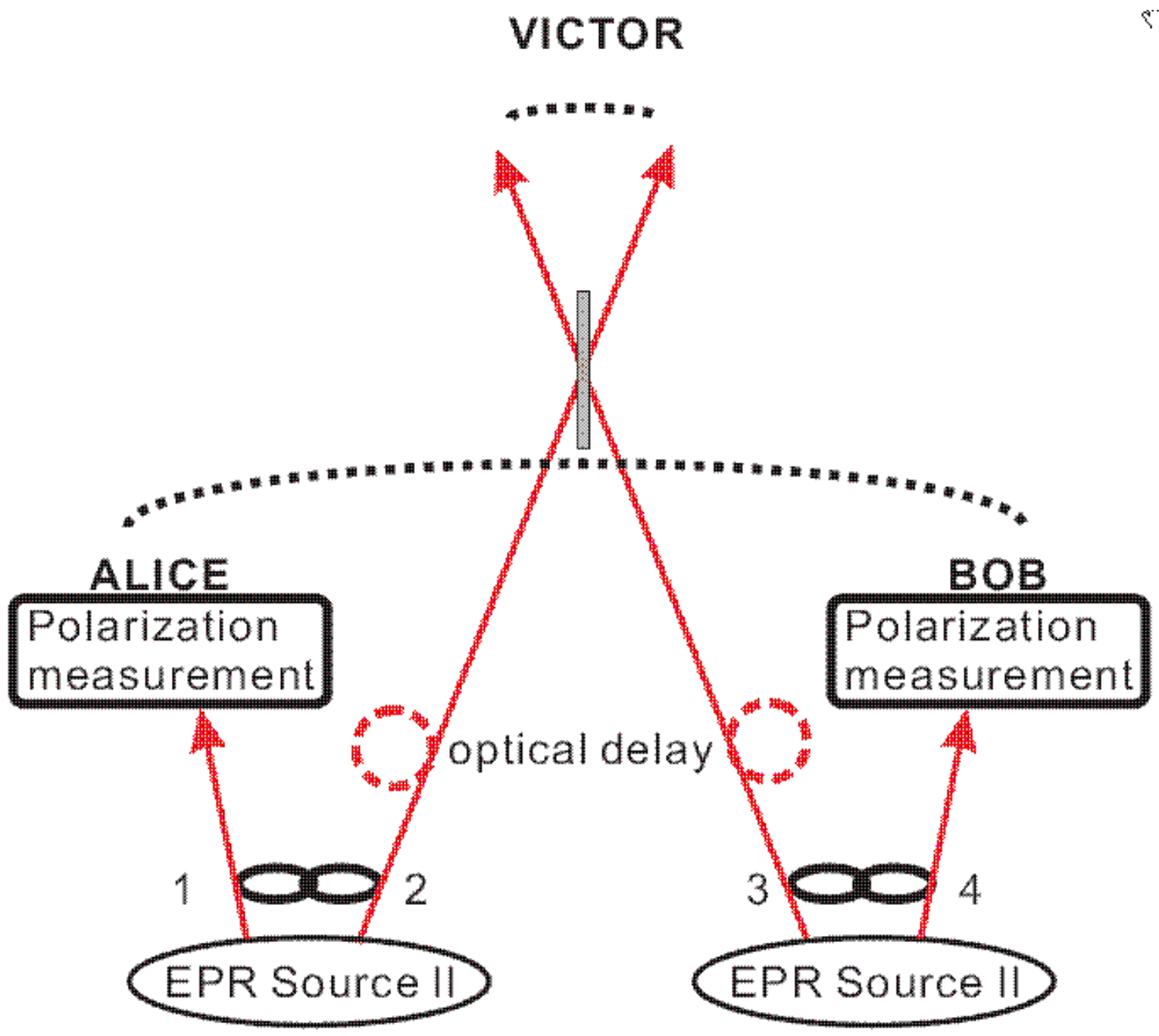

Entanglement Swapping

Consider any such four-fold detection. The initial quantum state has the form of a product of two EPR states $|\Psi^-\rangle_{12}|\Psi^-\rangle_{34}$. This may be expanded in terms of the four Bell states

$$|\Psi^-\rangle = 1/\sqrt{2}\ (\ |HV\rangle - |VH\rangle\ )$$
$$|\Psi^+\rangle = 1/\sqrt{2}\ (\ |HV\rangle + |VH\rangle\ )$$
$$|\Phi^-\rangle = 1/\sqrt{2}\ (\ |HH\rangle - |VV\rangle\ )$$
$$|\Phi^+\rangle = 1/\sqrt{2}\ (\ |HH\rangle + |VV\rangle\ )$$

as follows

$$|\Psi^-\rangle_{12}|\Psi^-\rangle_{34} = 1/2(|\Psi^+\rangle_{14}|\Psi^+\rangle_{23} - |\Psi^-\rangle_{14}|\Psi^-\rangle_{23} - |\Phi^+\rangle_{14}|\Phi^+\rangle_{23} + |\Phi^-\rangle_{14}|\Phi^-\rangle_{23}) \qquad (5)$$

Analysis of the actual experimental setup shows that cases (*i*) in which Victor records one photon as detected to each side of the beam splitter (with the same polarization) have non-zero Born probability only from the fourth term in (5), while cases (*ii*) in which Victor records both photons as detected to the same side of the beam splitter (with opposite polarizations) have non-zero Born probability only from the third term in (5).

What polarization quantum state should Victor ascribe to a pair 1+4? In a case (*i*) he should ascribe the corresponding 1+4 pair the quantum state $|\Phi^-\rangle_{14}$, while in a case (*ii*) he should ascribe the corresponding pair 1+4 the state $|\Phi^+\rangle_{14}$. In either case, Victor should ascribe an entangled state to systems that have never interacted, directly or indirectly. Moreover, in either case Victor ascribes a quantum state to a pair of systems *after* each system has been detected and no longer has any independent existence. The function of such quantum state ascriptions is perfectly standard. By inserting the relevant quantum state into the Born rule, any agent in Victor's situation can adjust its expectations concerning matters of which it is currently ignorant, namely what is recorded by Alice and Bob's detectors. Such expectations can be (and in the actual experiment were) compared to Alice and Bob's records in many cases of type (*i*) and many cases of type (*ii*). Those records returned statistics in conformity to the Born rule and in violation of Bell inequalities (as the polarization axes *a*, *b* were suitably varied). What this experiment illustrates in a striking way is that an agent may need to form expectations concerning events that have already happened although his physical situation renders him inevitably ignorant of their outcomes. Until physical interactions have suitably correlated Victor's records with the records of Alice and Bob's detectors and their environment, the objective Born probabilities he derives from the quantum state for his agent situation remain his most reliable guide to belief about their records.

There is a further feature of this experiment that involves delayed choice. Clearly, if the beam splitter were not present there would be no swapping of entanglement: 1+2 would remain entangled, as would 3+4, but there would be no entanglement across these pairs. The experiment is therefore designed to allow a "decision", effectively as to whether or not to introduce the beam splitter, to be postponed until *after* photons 1 and 4 have been detected. (In the actual experiment, implementing the "decision" was a little more complex, and was carried out by a quantum random number generator rather than any agent, human or otherwise.) With this additional feature, whether an agent in Victor's situation ascribes an entangled or a separable state to photons 1 and 4 depends on events that occur after these photons have been detected. The delayed-choice entanglement-swapping experiment reinforces the lesson that quantum states are neither descriptions nor representations of physical reality. In particular, it undermines the idea that ascribing an entangled state to quantum systems is a way of representing some new, non-classical, physical relation between them. To hold onto that idea in the context of this experiment would require one to maintain not only that *which* entanglement relation obtains between a pair of photons at some time, but also whether *any* such relation then obtains between them, depends on what happens to other independent systems later, after the pair has been absorbed into the environment.

Note also that the Born probabilities flowing from these assignments of quantum states cannot here be understood as chances that conform to Lewis's Principal Principle. That principle is supposed to explicate the role of chance in decision-making at a time by saying how an agent should base his credence of an event on the event's chance at that time. We saw earlier that to be guided by quantum theory in his decision-making, an agent may need to take account of information to which he is privy but that must remain inaccessible to other agents at that time. We now see a situation in which quantum theory supplies objective probabilities concerning a pair of events to no agent until after those events have occurred. Clearly these probabilities cannot guide any agent in forming credences about these events ahead of time. But they can still be useful to a decision-maker like Victor who is kept in ignorance of Alice and Bob's results.

Such ignorance would be irremediable in a modified scenario where the choice and Victor's measurements are *space-like* separated from Alice's and Bob's measurements (assuming no-superluminal signaling). In Lewis's terminology, that scenario would render information about Alice's and Bob's results inadmissible for Victor for a while *after* they had already occurred in his reference frame. Even thought they are objective, Born probabilities are indexed not to a time, but to the physical situation of a potential agent relative to the events they concern. These probabilities sometimes, but not always, depend on the temporal aspect of this relation. But whether or not they do, they may depend also on further physical aspects.

**5. The objectivity of physical description in quantum theory.** If quantum state ascriptions and the consequent Born probabilities are relative to agent situations, then is there any non-relational physical description on which agents in all situations can agree? The analogous question in the context of relativity theory receives a straightforward answer: frame-dependent descriptions including those of length and time-intervals may be thought to derive from frame-independent invariants such as the space-time interval. The question must be answered positively in the context of quantum theory in order to facilitate descriptive claims about the physical world any agent can endorse whatever that agent's physical situation, so that these claims can contribute to the predictive and explanatory goals of physics.

Section 3 explained why no claim expressed by a sentence of the form *S*: "The value of A on s lies in Δ" (no NQMC) is implied by any quantum state ascription (QSA) or Born probability statement (BPS)—even in a case in which *S* is correctly assigned Born probability 1 (relative to some agent situation). NQMCs were frequently and correctly made before the development of quantum theory and continue to be made after its widespread acceptance, which is why I call them non-quantum. While quantum theory adds no new ways of describing the physical world, it does offer authoritative advice to a situated agent, both on the content of NQMCs relevant to its situation, and on the degree of belief appropriate to such claims. The appropriate degree of belief generally depends on the agent situation, so differently situated agents are frequently advised to hold different epistemic attitudes toward NQMCs. But the content of a NQMC about a system *s* does not depend on agent situation. That is why any NQMC can be taken to offer a physical description that is objective, in the sense that the content of the claim is strongly agent-independent: it is independent of the physical as well as the epistemic state of any agent (human, conscious, or neither) that may make or evaluate it. But acceptance of quantum theory so drastically limits the content of *some* NQMCs that they can no longer contribute to the explanatory or predictive goals of physics and so are best left unsaid.

The distinctions just drawn between quantum state ascriptions, Born probability statements and non-quantum magnitude claims are important in explaining why quantum violation of Bell inequalities involves no physically problematic non-locality. Recall the discussion of non-local correlations in section 4. Remember that the event $M_B$ when Bob measures the polarization of photon *R* along the *b*-axis occurs earlier in the laboratory frame than the event $M_A$ when Alice measures the polarization of photon *L* along the *a*-axis. How should the relevant systems be described by NQMCs? Consider the following claims:

$D_1$: Photon-detector *L* records polarization *a* or polarization $a^{\perp}$.

$D_2$: Photon-detector *L* records polarization *a.*

$P_1$: Photon *L* has polarization *b* or polarization $b^{\perp}$.

$P_2$: Photon *L* has polarization *b*.

Assume (all experts including Alice and Bob agree that) a photon pair is emitted in polarization state $|\Phi^{+}\rangle$ and detection is perfectly efficient. $D_1$ can be taken to express an NQMC that is licensed by quantum theory (because of the decohering polarization-correlation interaction between *L* and its detector+environment) and warranted for any agent at any time, simply because Alice's photon detector is in good working order. $D_2$ can similarly be taken to express

an NQMC that is licensed by quantum theory, but whereas $D_2$ is warranted for Alice as soon as she records the *L* polarization, Bob is justified in making claim $D_2$ only when Alice communicates to him the record of the *L* polarization later---unless $b=a$, in which case Bob will be justified in claiming $D_2$ as soon as his detector records polarization *a* on photon *R*. (If $b \neq a$ then $D_2$ will be at most partially warranted for Bob before receipt of Alice's message.)

Now consider claims $P_1$ and $P_2$. One might suppose that Bob is justified in claim $P_2$ after consulting his photon detector *R* and finding that it has recorded polarization *b*, and that his entitlement extends to claim $P_1$ by simple logic. But this is incorrect. Even if $M_B$ occurs invariantly earlier than $M_A$ , quantum theory grants an agent (like Bob) only an extremely limited license to claims $P_1$ and $P_2$: the content of these claims is severely weakened by the severe restrictions on what material inferences they support. This is because in the interval between $M_B$ and $M_A$ photon *L* undergoes no interaction with its environment capable of delocalizing its quantum polarization state in a preferred $b$-$b^{\perp}$ basis. After his detector records polarization *b* for *R*, Bob's situation does warrant him in ascribing polarization state $|b\rangle$ to photon *L*: this becomes his new quantum polarization state for that photon. But no QSA implies any NQMC: the eigenstate-eigenvalue link fails. $P_2$ does support some material inferences to behavior characteristic of classically polarized light, importantly including that photon-detector *L* would record polarization *b*, and indeed will do so if $b=a$. But Bob is already warranted in believing these conclusions by his assignment of quantum state $|b\rangle$ to photon *L*, and so is any agent (including Alice) who knows the result of Bob's measurement. Any agent can reach these conclusions by assigning $|\Phi^{+}\rangle$ to the photon pair and conditionalizing on Bob's recorded *R* polarization *b*. So this part of the content of claim $P_2$ can be secured without it. Moreover, $P_2$ will not support additional material inferences just because *L* undergoes no decoherence interaction robustly correlating its *b*-polarization state to an environment. These considerations also undercut claim $P_1$. So Bob has no warrant for claiming $P_2$ if the content of this claim is taken to extend beyond that of the claim that his quantum polarization state for *L* is $|b\rangle$, and no warrant for claiming $P_1$ if the content of that claim is taken to extend beyond that of the claim that his quantum polarization state for *L* is either $|b\rangle$ or $|b^{\perp}\rangle$. These claims have no place in a careful agent-independent physical description capable of explaining the non-local correlations. Nor do analogous claims about the polarization of *R*.

Notice that we have now provided a reason why quantum theoretical analysis of these non-local correlations does not involve attributing to the polarizations of the *L*, *R* photons in a pair

"…any mutually independent existence (state of reality) [when] viewed separately, *not even if [they] are spatially separated from one another at the particular time under consideration*."[22]

At most, quantum theory licenses a claim about the polarization prior to detection of the *L*-photon only with respect to the *b*-axis, and a claim about the polarization prior to detection of the *R*-photon only with respect to the *a*-axis: and the license it extends to *these* claims is so severely restricted that neither amounts to a report on an independently existing state of reality. This is fortunate, since a well-trodden path takes one from the independent existence of pre-existing polarization states of both photons along every axis that their respective detectors simply reveal to Bell inequalities, whose experimentally-confirmed violation confirms quantum theoretical predictions derived from the Born rule as applied to state $|\Phi^{+}\rangle$. If each arbitrarily-oriented photon-detector faithfully revealed a pre-existing polarization of the detected photon, and these polarizations were distributed among many pairs in state $|\Phi^{+}\rangle$ in a way that was independent of the detector settings *a*, *b*, then the only way to restore consistency with the Born rule predictions would be to allow that the polarization state of a photon could be non-locally altered before reaching the detector. That would constitute a blatant violation of a physical locality condition Einstein(1948, 322) stated as follows

> aussere Beeinflussung von A hat keinen *unmittelbaren* Einfluss auf B; dies ist als << Prinzip der Nahewirkung >> bekannt ("an external influence on A has no immediate effect on B; this is known as the 'principle of local action' ")

[22] Einstein (1949, 681-2)

The present pragmatist approach to quantum theory acquits it of any such violation of local action.

Bell inequalities may be derived in a bipartite system like the photon pair *LR* without assuming anything corresponding to independently existing polarizations for each subsystem. The key assumptions here are of conditional probabilistic independence such as the following (Gisin, 2009) where I suppose that α, β are variables ranging over the possible values of polarization recorded by *L, R* detectors respectively along the *a*, *b* axes, and λ (which may include the quantum state, here $|\Phi^{+}\rangle$) specifies the situation of everything in the past irrelevant to the choice of *a*, *b* axes:

$$\text{prob}(\alpha|a, b, \lambda) = \text{prob}(\alpha|a, \lambda) : \text{prob}(\beta|a, b, \alpha, \lambda) = \text{prob}(\beta|b, \lambda) \quad (6)$$

Which together imply

$$\text{prob}(\alpha, \beta|a, b, \lambda) = \text{prob}(\alpha|a, \lambda) \times \text{prob}(\beta|b, \lambda) \quad (7)$$

Gisin glosses these conditions as follows

> ...for any give "state of affairs" λ, what happens on Alice's side does not depend on what happens on Bob's side, and vice versa.

But if λ is just whatever earlier physical conditions warrant ascription of quantum state $|\Phi^{+}\rangle$ to the pair, and the probabilities appearing in (6) are taken to be consequences of the Born rule as applied to $|\Phi^{+}\rangle$, then neither (6) nor (7) says anything about what happens in this situation. As in the BPSs from which they follow, (6) and (7) describe nothing in the physical world: their role is simply to offer authoritative advice to an agent such as Alice or Bob on what to expect in the situation described. As section 2 explained, a Born probability statement does not purport to describe physical reality. Its role within the theory is to offer objective advice to a physically situated agent on how to apportion beliefs concerning matters of which it is ignorant. So understood, conditions like (6) and (7) do not express physical locality conditions. The fact that Born probabilities violate the second part of (6) does not make the quantum world physically non-local. But it is interesting to note that if Born probabilities had violated just the *first* part of (6), by choosing one axis of his detector rather than another during repeated runs of the experiment, Bob would have been able to guide Alice's expectations and thereby manipulate her behavior in a way that *would* have violated Einstein's principle of local action! So the no-signaling theorems remain critical to this acquittal of quantum theory from the charge of violating a physically motivated locality condition.

The content of a NQMC expressed by $D_1$ or $D_2$ is in no way relative to the situation of any actual or possible agent, even though an agent's situation may well affect its warrant for making that claim. Moreover, because of the nature of the decohering polarization-correlation interaction between *L* and its detector+environment, the inferential power of one of these claims extends very far—far enough for such claims to be considered simply objective descriptions of the physical world for all practical and impractical purposes of any agent. An agent making such a claim may therefore be understood to be offering an objective description of the physical event normally taken to constitute the outcome of a measurement of linear polarization of $L$ along the $a$ axis. It is by licensing, though not implying, such claims that quantum theory authorizes objective physical description of the world.

As we have seen, the status of NQMCs $P_1$ and $P_2$ is different. Even though their content need not be regarded as relative to the situation of any agent making them, quantum theory severely limits their inferential power. This so restricts their content that it is no longer appropriate to think of these claims as simply offering an objective description of the physical properties of photon $L$, or of any other physical state of affairs. This does not mean that such claims have no use. Bob may utter $P_2$ intending thereby merely to ascribe quantum state $|b\rangle$ to photon $L$, and such a usage may acquire common currency within a community of quantum physicists. Indeed, physicists do often ascribe linear polarization states to photons. But this does not show that by saying something like $P_2$ these physicists are offering objective physical descriptions. By familiarizing themselves with quantum theory they have internalized the limited inferential power attached to such a claim and they use it with that common understanding.

The "paradox" of Wigner's friend presents a challenge to the objectivity of physical description within quantum theory. To set up the "paradox", imagine Schrödinger's cat (and associated "diabolical device") replaced by a human experimenter (Wigner's friend) who records in a device $D$ the result of a quantum measurement he has performed on a system $s$ inside his isolated laboratory.[23] For example, suppose that $s$ is the photon $L$ just considered, $D_L$ detects its polarization along the $a$ axis, and the friend is Bob. Meanwhile, another agent $W$ remains outside the laboratory. After consulting the record of $D_R$, Bob ascribes quantum polarization state $|b\rangle$ to $L$ and applies the Born rule to calculate the probability, $(|\langle a|b\rangle|^2)$, of $D_2$. He performs the measurement, consults $D_L$, and records the outcome by writing $D_2$ in his notebook. $W$, on the other hand, ascribes a quantum state $|\Psi_W\rangle$ to the enormously complex system composed of $L$, $R$, Bob, $D_L$, $D_R$, the notebook and everything else in the laboratory, and uses this to calculate the Born probability of $D_2$. Even though this will not equal $(|\langle a|b\rangle|^2)$, no inconsistency arises, since all BPSs are relative to the physical situation of an agent making them, and $W$ and Bob are in relevantly different physical situations: at this stage Bob has interacted, first with $D_R$, and then with $D_L$, but $W$ has interacted with neither.

An inconsistency would arise if one took a quantum state completely to describe a system in accordance with the eigenstate-eigenvalue link and assumed measurement collapses this state onto an eigenstate corresponding to the corresponding eigenvalue of the measured observable. Bob would then take his measurement of $L$-photon polarization to collapse the state of $L$-photon+$D_L$ onto an eigenstate of which a "pointer reading" on $D_L$ was an eigenvalue. But, treating Bob merely as part of the physical contents of the laboratory, $W$ would *deny* that the state $|\Psi_W\rangle$ collapsed onto such an eigenstate until *he*, $W$, made a measurement by entering the laboratory to see what polarization $D_L$ had recorded. There is no threat of such inconsistent descriptions of the contents of the laboratory prior to *W*'s entry on the present pragmatist approach, which denies any descriptive role to quantum states. But if quantum theory denies *W* the license to use his quantum state $|\Psi_W\rangle$ to describe what is happening in the laboratory before he enters, while licensing Bob to make descriptive claims such as $D_2$, then how can a claim like $D_2$ be taken to offer an objective physical description ?

---

[23] To call the laboratory 'isolated', is to require by *fiat* the absence of any decohering interactions with its external environment. So we are talking of a ridiculously impractical *Gedankenexperiment*, as Schrödinger explicitly said he was when describing his cat scenario. The point of doing so is to explain why even such an extremely hypothetical situation would pose no threat to the objectivity of physical description in quantum theory.

A default assumption underlying the objectivity of physical description dictates that *W* accept Bob's sincere report $D_2$ when backed up by *W*'s own independent investigations. This assumption is so deeply embedded in scientific methodology that it is hard to imagine how any kind of scientific activity could survive its wholesale rejection. *W*'s quantum analysis of his situation may seem to challenge this assumption. Since he knows that Bob was to prepare his photon pair in polarization state $|\Phi^+\rangle$, *W*'s initial quantum state $|\Psi_W\rangle_{t1}$ will include a representation of the polarization state of the *L*-photon that is a superposition of $|a\rangle$ and $|a^\perp\rangle$ with appropriate non-zero coefficients. Subsequent interactions with $D_L$, further recording equipment, Bob and Bob's notebook will entangle this superposition with their quantum states. Nothing about this quantum state will even suggest what result (if any) Bob got in his measurement of the polarization of the *L*-photon. But *W*'s quantum state will advise him to expect that, whatever that result may be, the laboratory will contain multiple mutually supporting records of it. So that while his quantum analysis alone provides *W* no warrant for believing $D_2$, it does warrant *W* in believing that his observation of $D_L$, consultation with Bob, reading Bob's notebook, and any other examination of what is ordinarily taken to be evidence that $D_2$ was true even before he entered the laboratory, will all be mutually consistent with each other, and consistent also—*either* with $D_2$ *or else* with $D_2{}^\perp$: Photon-detector $D_L$ records polarization $a^\perp$. It remains perfectly consistent with *W*'s quantum analysis of the situation for him to suppose that it is $D_2{}^\perp$, not $D_2$, that correctly described the physical situation prior to his entering the laboratory, despite Bob's sincere statement that he remembers recording $D_2$ , backed up by all *W*'s own observations on entering the laboratory.

There is nothing strictly paradoxical about this situation. But it does prompt the skeptical concern that an agent who accepts quantum theory no longer has any reason to expect apparently sincere reports of fellow agents concerning readily observable properties of macroscopic objects to be reliable, not even if these are backed up by its own independent observations of these properties, together with what are ordinarily considered traces of them. Quantum theory does not validate the default assumption underlying the objectivity of physical description.

I think the right way for a scientist to respond to this concern is simply to refuse to take this skeptical possibility seriously. A scientist begins by trusting his or her own observations as well as those of others and questions these only when further observations provide positive reasons for doing so. Nothing we learn from quantum theory or anything else in science provides *W* with any reason for questioning his own or his friend's sincere observation reports concerning the outcomes of quantum measurements or the gross properties of macroscopic objects. Science concerns itself precisely with those physical descriptions that can be taken to be objective in the sense that they are open to support from multiple independent observations whose evidential import can be collectively undercut only by this kind of radical philosophical skepticism, yielding to which would render scientific investigation of any kind impossible. The extremely hypothetical scenario of Wigner's friend fails to lift the burden of proof from one who would seek to deny the objectivity of physical descriptions such as that offered by the claim $D_2$.

A further twist on the Wigner's friend scenario will help to bring out a quantum limitation on the content of all NQMCs (including not only $D_2$ but also claims such as $S_x$ from section 3), and indeed on all physical descriptions. Consider *W*'s quantum state $|\Psi_W\rangle$. Since the entire laboratory and its contents constitutes an isolated system, *W* will take $|\Psi_W\rangle$ to have evolved unitarily from its state $|\Psi_W\rangle_{t1}$ prior to his friend's measurement of the polarization of *L* to its state $|\Psi_W\rangle_{t2}$ just as he enters the laboratory to ask his friend about its result.

$$|\Psi_W\rangle_{t2} = U_{12} |\Psi_W\rangle_{t1} \qquad (8)$$

*W* should ascribe to the contents of the laboratory at $t_1$ a quantum state that reflects his belief that his friend has not yet performed the planned measurement on *L*. So *W* will be warranted in ascribing to these contents a state $|\Psi_W\rangle_{t1}$ that assigns Born probability 1 to NQMCs on $D_L$, Bob and his notebook that suffices to substantiate that belief. Mathematically, there will exist a Hamiltonian that would induce the time-reversed evolution of $|\Psi_W\rangle$ so that at a later time $t_3$ (where $t_3 - t_2 = t_2 - t_1$) it is restored to its value before the friend measured the polarization of *L*

$$|\Psi_W\rangle_{t3} = U^{\dagger}_{23} |\Psi_W\rangle_{t2} = |\Psi_W\rangle_{t1} \qquad (9)$$

If *W* had the powers of a quantum demon, he could instantaneously replace the original Hamiltonian by this time-reversing Hamiltonian at $t_2$, thereby restoring $|\Psi_W\rangle$ at $t_3$ to its original value at $t_1$.[24] Suppose that he does so, and postpones his entry into the laboratory until $t_3$. Since the quantum state of the entire laboratory is identical to what it was before his friend had made any measurement of the polarization of *L*, *W* must fully expect that if he then asks his friend about the result of his measurement, the friend will say he has not yet performed any measurement. He must further fully expect that his own examination at, and at any time after, $t_3$ of $D_L$, Bob's notebook, and anything else inside the laboratory will reveal no record of any such measurement ever having been made. *W*'s action at $t_2$ has, by $t_3$, erased all traces of Bob's measurement of the polarization of *L* and its result: Indeed, *W* has succeeded in erasing all traces of everything that happened inside the laboratory between $t_1$ and $t_3$.

It is deeply embedded in the way we ordinarily think about the past that everything that happens leaves some trace of its occurrence, however epistemically inaccessible this may be to us. Dummett(1969) even took rejection of this assumption to be a significant motive for antirealism about the past---the view that statements about the past on which no present or future evidence bears have no determinate truth-value. Consideration of the extended Wigner's friend scenario shows that one who wholly accepts quantum theory must limit the content of NQMCs and indeed all other physical descriptions so that such a claim does not thereby exclude the physical possibility that the claimed state of affairs leave no trace whatever. But while allowing for this possibility does marginally weaken every physical description I do not see that quantum theory thereby makes physical description any the less objective.

---

[24] While completely out of the question for such a complex system, such reversals have been seriously considered as a way to restore coherence in a mesoscopic system consisting of an electron on a quantum dot interacting with about a million nuclear spins (see Yao *et al.*(2007)).

**6. Conclusion.** I promised a diagnosis of the curious situation that while there are no serious and lasting disagreements among physicists on the use of quantum theory, disputes about its meaning began with its inception and continue unchecked. I suggest that what is largely responsible for this situation is the assumption (tacit or explicit) that the meaning of quantum theory must be given by saying what the physical world is like, according to that theory. That the meaning of *any* theory is to be given by spelling out its truth-conditions is an assumption no pragmatist will leave uncontested. Rejecting this blinkered perspective on interpretation makes it possible to see that the real significance of quantum theory for the philosophy of science is how it advances the goals of physics *without* presenting us with novel ways of representing the world. In this paper I have indicated how this new perspective permits progress on long-standing problems such as the measurement problem and quantum non-locality: many details remain to be filled in subsequently.

One goal of physics emphasized by realists is explanation of natural phenomena. I anticipate the objection that quantum theory is only able to achieve its great explanatory success through is deployment of novel theoretical representations of the world. Since this pragmatist interpretation cannot account for quantum theory's explanatory successes (the objection continues), it is merely a new name for a bad old way of thinking–instrumentalism! To confine the present paper within reasonable bounds, a second paper will be devoted to a detailed refutation of this objection.